\documentclass[apj]{emulateapj}
\usepackage{graphicx,color}
\usepackage{float}
\usepackage{amsmath}
\usepackage{natbib}
\usepackage{subfigure}
\usepackage[breaklinks]{hyperref}
\bibpunct{(}{)}{;}{a}{}{,}

\begin{document}
% Title Page

\submitted{Published 2012 March 15}
\title{The relationship between hard X-ray pulse timings and the locations of footpoint sources during solar flares}
%\titlerunning{Pulse timings and locations during solar flares}

%\institute{Solar Physics Laboratory, Heliophysics Science Division, NASA Goddard Space Flight Center, Greenbelt, MD, 20771, USA}

%\author{A. R. Inglis
%\and B. R. Dennis} 

\author{A. R. Inglis and B. R. Dennis}
\affil{Solar Physics Laboratory, Heliophysics Science Division, NASA Goddard Space Flight Center, Greenbelt, MD, 20771, USA}

%\abstract {}{To investigate quasi-periodic pulsations (QPP) observed during the flare of 9th November 2002.}{Some preliminary results while studying this flare using GOES and %RHESSI.}{It looks like there are QPPs.}{}

\begin{abstract}
 The cause of quasi-periodic pulsations (QPP) in solar flares remains the subject of debate. Recently, \citet{2011ApJ...730L..27N} proposed a new model suggesting that, in two-ribbon flares, such pulsations could be explained by propagating slow waves. These waves may travel obliquely to the magnetic field, reflect in the chromosphere and constructively interfere at a spatially separate site in the corona, leading to quasi-periodic reconnection events progressing along the flaring arcade. Such a slow wave regime would have certain observational characteristics. We search for evidence of this phenomenon during a selection of two-ribbon flares observed by RHESSI, SOHO and TRACE; the flares of 2002 November 9, 2005 January 19 and 2005 August 22. We were not able to observe a clear correlation between hard X-ray footpoint separations and pulse timings during these events. Also, the motion of hard X-ray footpoints is shown to be continuous within the observational error, whereas a discontinuous motion might be anticipated in the slow wave model. Finally, we find that for a preferential slow wave propagation angle of 25-28 degrees that is expected for the fastest waves, the velocities of the hard X-ray footpoints lead to estimated pulse periods and ribbon lengths significantly larger than the measured values. Hence, for the three events studied, we conclude that the observational characteristics cannot be easily explained via the  \citet{2011ApJ...730L..27N} propagating slow wave model when only angles of 25-28 degrees are considered. We provide suggested flare parameters to optimise future studies of this kind.
\end{abstract}

\maketitle

\section{Introduction}

The nature of quasi-periodic pulsations (QPP) in solar flares has become a well-studied topic over the last few decades, from early hard X-ray observations by \citet{1969ApJ...155L.117P} and various radio band observations \citep{1970SoPh...13..420C, 1983ApJ...271..376K} to modern, spatially resolved observations throughout the electromagnetic spectrum \citep{2001ApJ...562L.103A,Grechnev,Melnikov,2005ApJ...620L..67M,2005A&A...440L..59F,2008A&A...487.1147I,2008MNRAS.388.1899S,2009A&A...493..259I,2009ApJ...697..420K,2009SoPh..258...69Z,2010ApJ...708L..47N}. However, it has not yet been possible to definitively identify the underlying physical mechanism responsible for QPP. 

Two main interpretations have risen to the fore in the search to understand this phenomenon. The first of these is that QPP are a signature of magnetohydrodynamic (MHD) wave modes. MHD waves are known to be supported by solar plasmas \citep{1982SoPh...76..239E,1983SoPh...88..179E}, and various wave modes are capable of modulating physical plasma parameters and hence direct observables \citep[see][for a comprehensive review of the properties of coronal waves]{2005LRSP....2....3N}. Hence it follows that wave modes in flare sites could give rise to the quasi-oscillatory behaviour we often observe in flaring lightcurves.

The second main interpretation of flaring QPP is that they are a signature of periodic, or bursty magnetic reconnection. Reconnection is known to act as an energy release trigger in flare sites, hence it is a natural extension to consider that this can occur in a repetitive regime \citep[e.g.][]{2009A&A...494..329M,2009A&A...493..227M}. Moreover, in contrast to the basic CKSHP flare model, a realistic 3D flare model is such that the reconnection site physically moves as the flare progresses \citep[][for example]{2006ApJ...642.1177L}, rather than being a stationary X-point. Hence the emission from flares could be directly modulated as the reconnection site moves and the reconnection rate varies. An extensive review on current trends in flare studies, including a discussion of QPP may be found in \citet{2011SSRv..159...19F}, while \citet{2009SSRv..149..119N} present a review focusing on QPP in particular.

Differentiating between these two possible explanations has proven observationally difficult, despite many advances in instrumentation over the years. Moreover, the distinctions between quasi-periodic pulsations and genuinely periodic oscillations is often blurred. In the first case we typically observe a series of emission peaks which show some evidence that they are temporally related, rather than the clearly periodic signal expected in the second case. In truth, very few observations fall into the second category, while precise criteria for the first category remain elusive.  For example, \citet{2011A&A...533A..61G} recently showed that some previous detections of QPP may be considered unreliable due to the employed detection methodology \citep[see also][]{2011A&A...530A..47I}. Hence care must be taken during the analysis of quasi-periodic events.

The most-used instruments for such studies currently in operation include the Reuven Ramaty High Energy Solar Spectroscopic Imager (RHESSI) \citep{2002SoPh..210....3L}, the Nobeyama Radioheliograph (NoRH)\citep{1994IEEEP..82..705N} and the Nancay Radioheliograph (NRH), while satellites such as SOHO, TRACE and the Solar Dynamics Observatory (SDO) are used to provide supporting imagery in the EUV range. RHESSI and NoRH in particular are heavily used because of their excellent time cadence, ideal for studying short, transitory phenomena, and in the case of RHESSI also its fine spatial resolution and wide coverage of the X-ray regime. SDO serves as an ideal supporting instrument because of its Atmospheric Imaging Assembly (AIA), which provides full-disk images of the Sun at multiple EUV wavelengths with a 12 s time cadence. However, due to solar minimum few two-ribbon flares have been observed since the launch of SDO. Therefore this study focuses on observations made with the SOHO and TRACE instruments. 

\begin{figure}[H]
 \begin{center}
  \includegraphics[width=8cm]{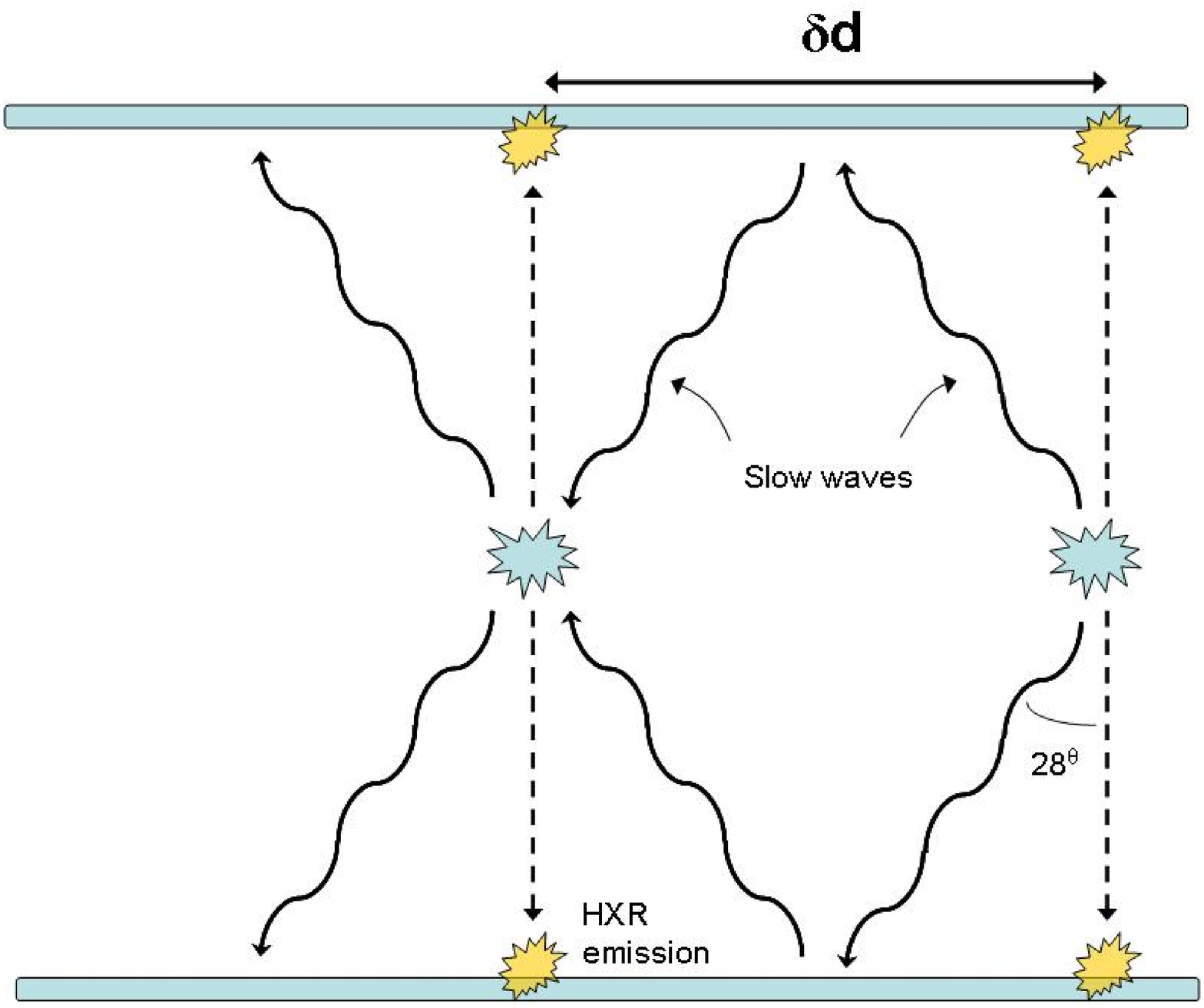}%{paper_cartoon2.ps}
  \caption{Two-dimensional cartoon of a two-ribbon arcade. Slow waves (solid arrows), excited by a reconnection event at some location in the flaring arcade, may propagate at an angle - preferentially 25-28 degrees for the fastest components - down the magnetic field to the footpoint regions where the ribbons - represented by the shaded areas - are located. The dense chromosphere causes these waves to reflect, and recombine at a point further along the arcade in the corona, triggering another reconnection event. Consequently, particles (dashed arrows) are accelerated at the reconnection sites and propagate directly along the magnetic field lines and encounter the thick-target chromosphere, thus generating hard X-ray emission. The distance between re-combinations is labelled $\delta d$. The total distance travelled by the slow wave is longer than $\delta d$, making the apparent velocity of the reconnection site and also the X-ray footpoints along the ribbon slower than the wave velocity. We denote the time between reconnection events as $\delta p$.}
  \label{cartoon}
 \end{center}

\end{figure}

Recently, \citet{2011ApJ...730L..27N} sought to address the question of whether MHD wave modes could explain pulsations in a complex 3D flare system, introducing a new model to explain pulsations in large, two-ribbon flares. In this model, it is realised that magnetoacoustic slow waves may propagate at an angle to the magnetic field, rather than directly along it, reflect from the chromosphere and act as triggers of reconnection when they constructively interfere at coronal sites (see Figure \ref{cartoon}). Hence a slow wave excited by a flare would propagate at an angle to the field down to the chromosphere, reflect and cause an additional burst of reconnection at a different location when these waves recombine in the corona. This leads to a progression of the reconnection region along a flaring arcade. Although the wave would diverge over a range of angles from the magnetic field, \citet{2011ApJ...730L..27N} showed that the preferential propagation angle - that is, the angle at which the wave travels with the highest velocity - is 25 - 28 degrees, and proposed that these fastest components are sufficient to trigger another energy release further along the arcade. Also, although typical slow wave speeds are of the order 100 - 300 km/s, this model means that the progression along the arcade occurs at a fraction of the slow wave velocity, broadly consistent with current observations of two-ribbon flares.

A recent follow-up numerical study by \citet{2011A&A...536A..68G} showed that the principle of this theory is sound, i.e. that the propagating slow waves would form a distribution about the magnetic field, as expected from the varying velocity as a function of propagation angle. This distribution is preserved as the waves are reflected at the chromosphere and return to the corona. However, although \citet{2011ApJ...730L..27N} cited some supporting examples, this model requires further observational testing, and provides the motivation for this work. We study three two-ribbon flares which exhibited QPP, and attempt to understand their nature based on their hard X-ray footpoint motion, as well as the correlation between these motions and the timing of the pulsations. For this initial study we consider only the fastest component of the slow wave, which travels at 25-28 degrees to the magnetic field as calculated by \citet{2011ApJ...730L..27N}.

\section{Event selection}

For this case study, we limit ourselves to three two-ribbon flares, each observed with at least one EUV imager. We require each event to be imaged by RHESSI in order to examine hard X-ray footpoint motions, and also require that the event consists of two clear footpoints in the hard X-ray range that move along the flare ribbons observed in the EUV range. We select three events of interest; the flares of 2002 November 9, 2005 January 19, and 2005 August 22, all of which exhibit QPP and have been the subject of previous studies \citep[e.g.][]{2005ApJ...625L.143G, 2009ApJ...693..132Y,2009SoPh..258...69Z,2008SoPh..247...77L,2010ApJ...724..171R,2011A&A...525A.112R}. These flares show a range of periods; the 2002 November 9 flare exhibits periods $P$ of = 30 - 80 s, whereas for the 2005 January 19 event $P$ = 50 - 200 s and for the 2005 August 22 flare $P$ = 100 - 250 s. This allows us to test for a slow wave regime over a range of scales. The flare information is summarised in Table \ref{table1}.

\section{Lightcurves}

\begin{figure}
 \begin{center}
  \includegraphics[width=8cm]{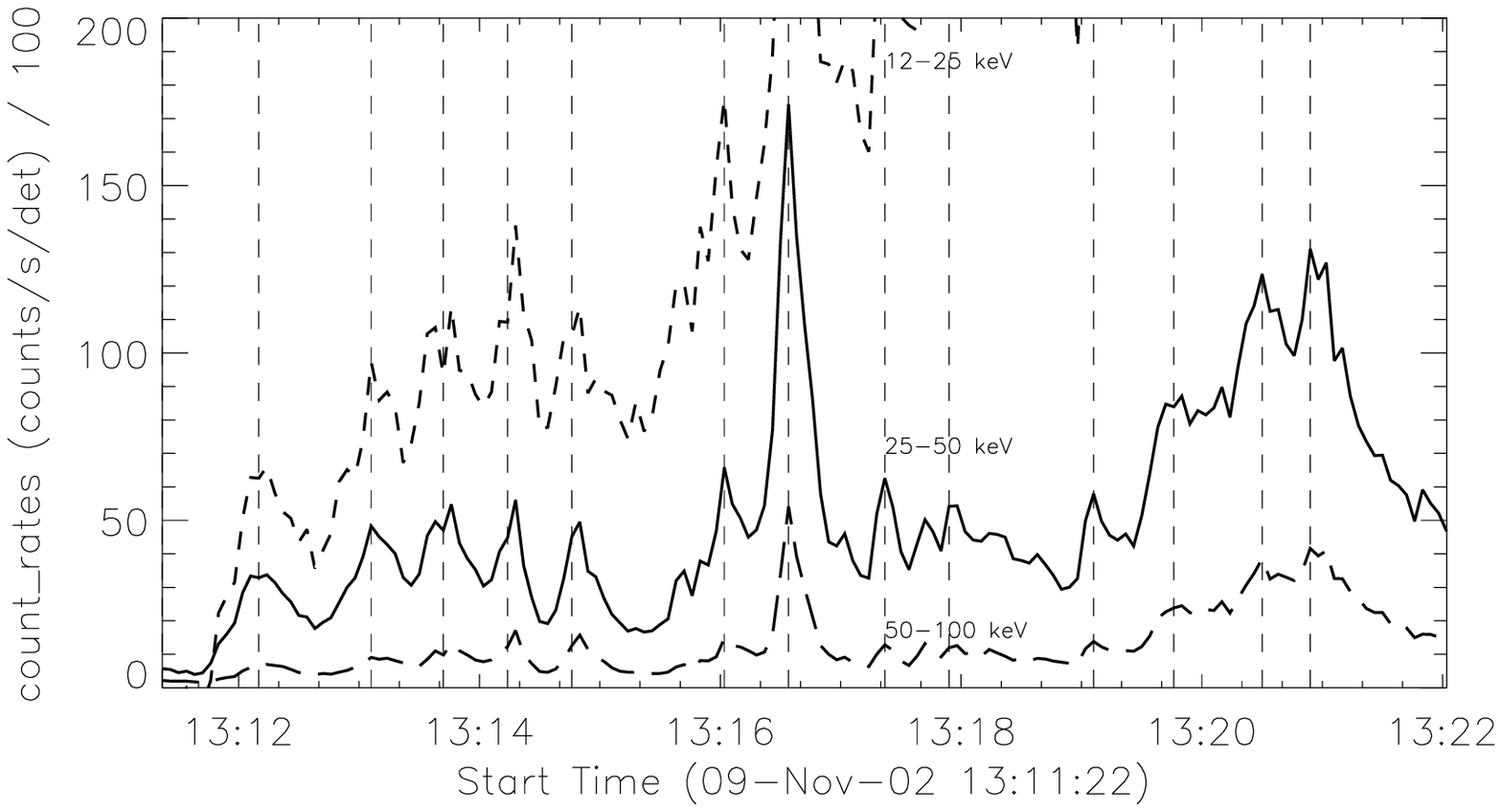}
   \includegraphics[width=8cm]{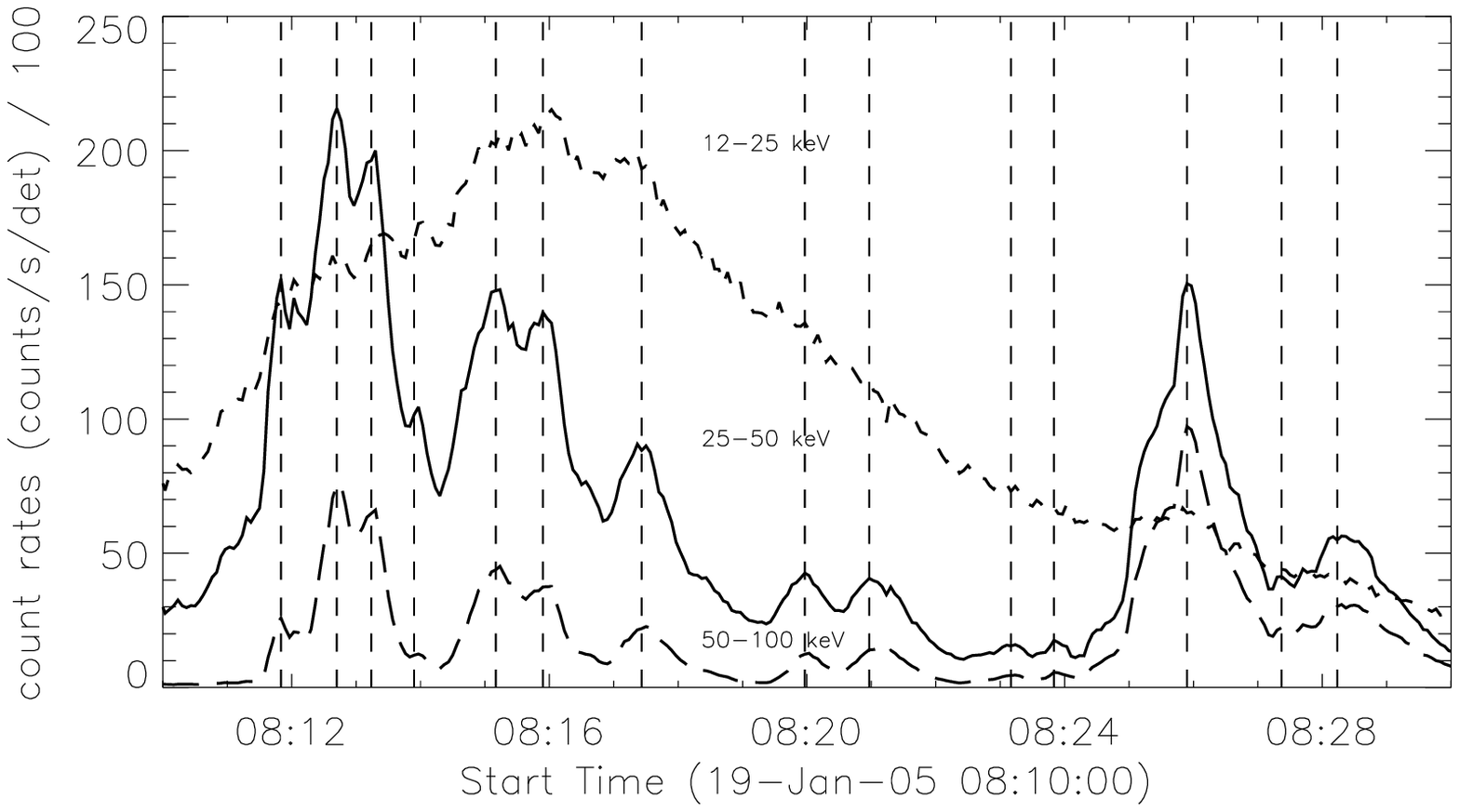}
   \includegraphics[width=8cm]{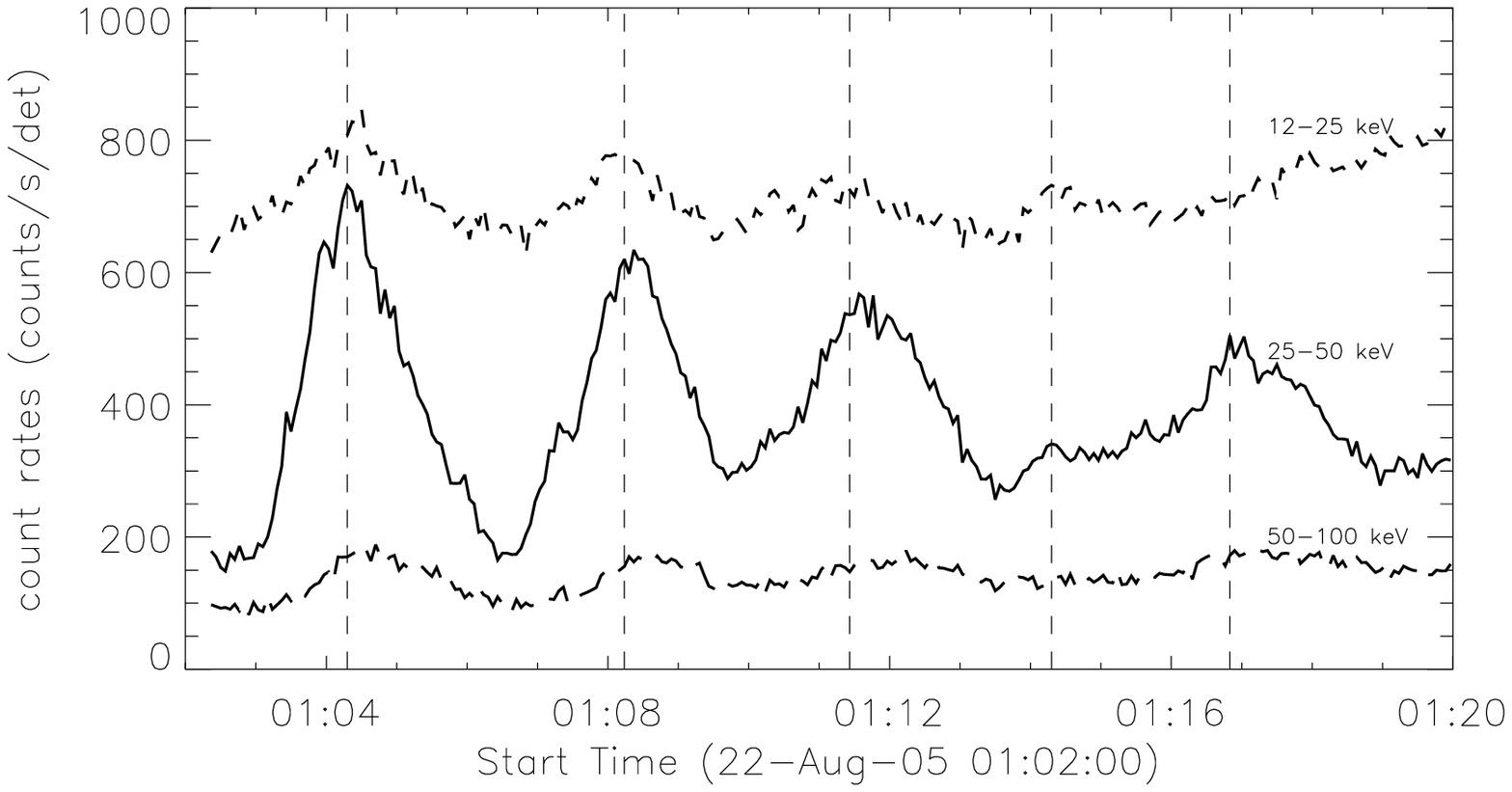}
\caption{Lightcurves of the selected two-ribbon flares exhibiting QPP, each observed by RHESSI in the 12-25 keV, 25 - 50 keV, and 50 - 100 keV energy bands. Top panel: the flare of 2002 November 9. Centre panel: the flare of 2005 January 19. Bottom panel: the flare of 2005 August 22. For clarity the 12 - 25 keV flux has been rescaled in each panel. The vertical lines indicate the pulses we consider, as described in Section \ref{source_positions}.}
\label{xray}
 \end{center}

\end{figure}

We investigate a flare of GOES-class M4.9 that began at 13:10 UT on 2002 November 9. This flare was previously investigated by \citet{2005ApJ...625L.143G} and \citet{2009ApJ...693..132Y}. It was observed by RHESSI, and quasi-periodic pulsations were evident in the RHESSI lightcurves. A selection of the energy ranges observed by RHESSI are shown in the first panel of Figure \ref{xray}. This flare was observed at multiple wavelengths by the Nancay Radioheliograph (NRH). Similar pulsations were observed both in these radio bands, and in the time derivative of the GOES flux, as anticipated by the Neupert effect. For brevity we do not reproduce these here.

%According to the Neupert effect, the derivative of soft X-ray emission should be similar to the observed hard X-ray emission. Thus, even though pulsations are not obvious in the %original GOES lightcurve, plotting their derivatives alongside RHESSI hard X-rays reveals the presence of similar pulsations. However, these pulsations do not appear to be %precisely in phase, particularly at 13:15:20 UT. The figure also shows that a state change occurs in the GOES data collection mode at approximately 13:19 UT, beyond which useful %data cannot be extracted due to a poor signal-to noise ratio.

The second flare of our study was an X1.3-class solar flare from 2005 January 19 starting at 08:00 UT. This flare was previously studied by \citet{2009SoPh..258...69Z} and \citet{2009ApJ...693..132Y}. A series of emission spikes are evident in the lightcurves over a 30 minute time interval, shown in the second panel of Figure \ref{xray}. This flare was cited by \citet{2011ApJ...730L..27N} as a possible example of the slow wave model due to the gradual motion of hard X-ray footpoints along the flare arcade at sub-sonic speeds; \citet{2009SoPh..258...69Z} measured the velocity of the northern footpoint at $v \approx$ 59 km/s. 

%\begin{figure}
% \begin{center}
 % \includegraphics[width=8cm]{19_jan_05_lightcurves.ps}
 % \caption{Lightcurves of the flare of 2005 January 19, observed with RHESSI in the 12-25 keV (dashed line), 25-50 keV (solid line) and 50 - 100 keV (dot-dashed line) bands.}
 %\end{center}

%\end{figure}

The flare of 2005 August 22 has been the subject of several previous studies, including \citet{2008SoPh..247...77L, 2009ApJ...693..132Y, 2010ApJ...724..171R,2011A&A...525A.112R}. The early stages of the flare were not observed by RHESSI as it was in the night-time phase of its orbit. This flare was observed in its entirety by the Nobeyama Radioheliograph (NoRH) in Japan, which showed that the flare onset was at 00:53 UT. RHESSI observations begin at 01:02 UT and continue past the end of the pulsations at 01:20 UT. During this time five pulses are observed in both the radio and X-ray lightcurves. It is noteworthy that in the hard X-ray channels the pulses are triangular in appearance.

\section{Spatial distribution of sources}
\label{source_positions}

\begin{figure}
 \begin{center}
    \includegraphics[width=8.3cm]{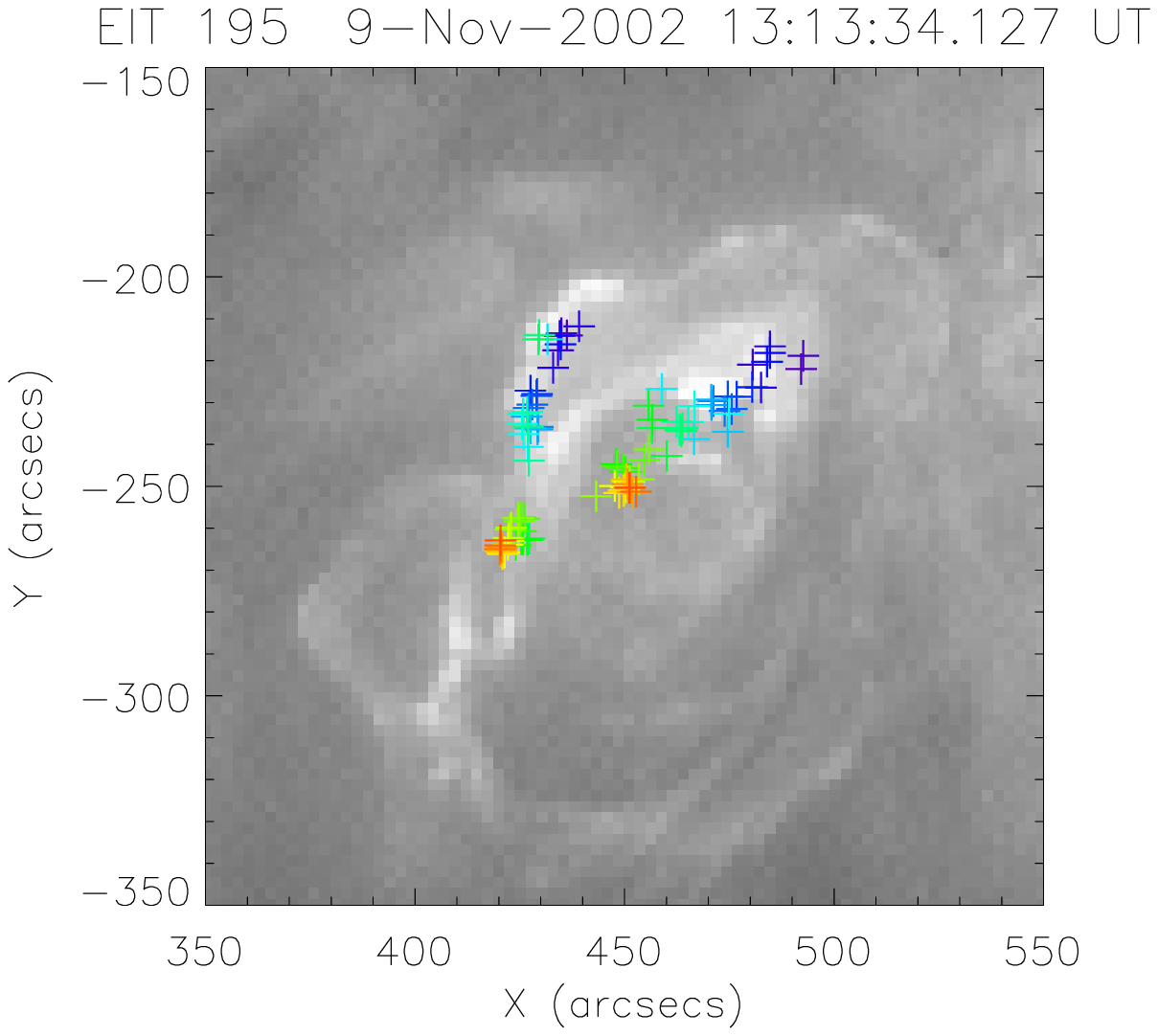}
    \includegraphics[width=8.3cm]{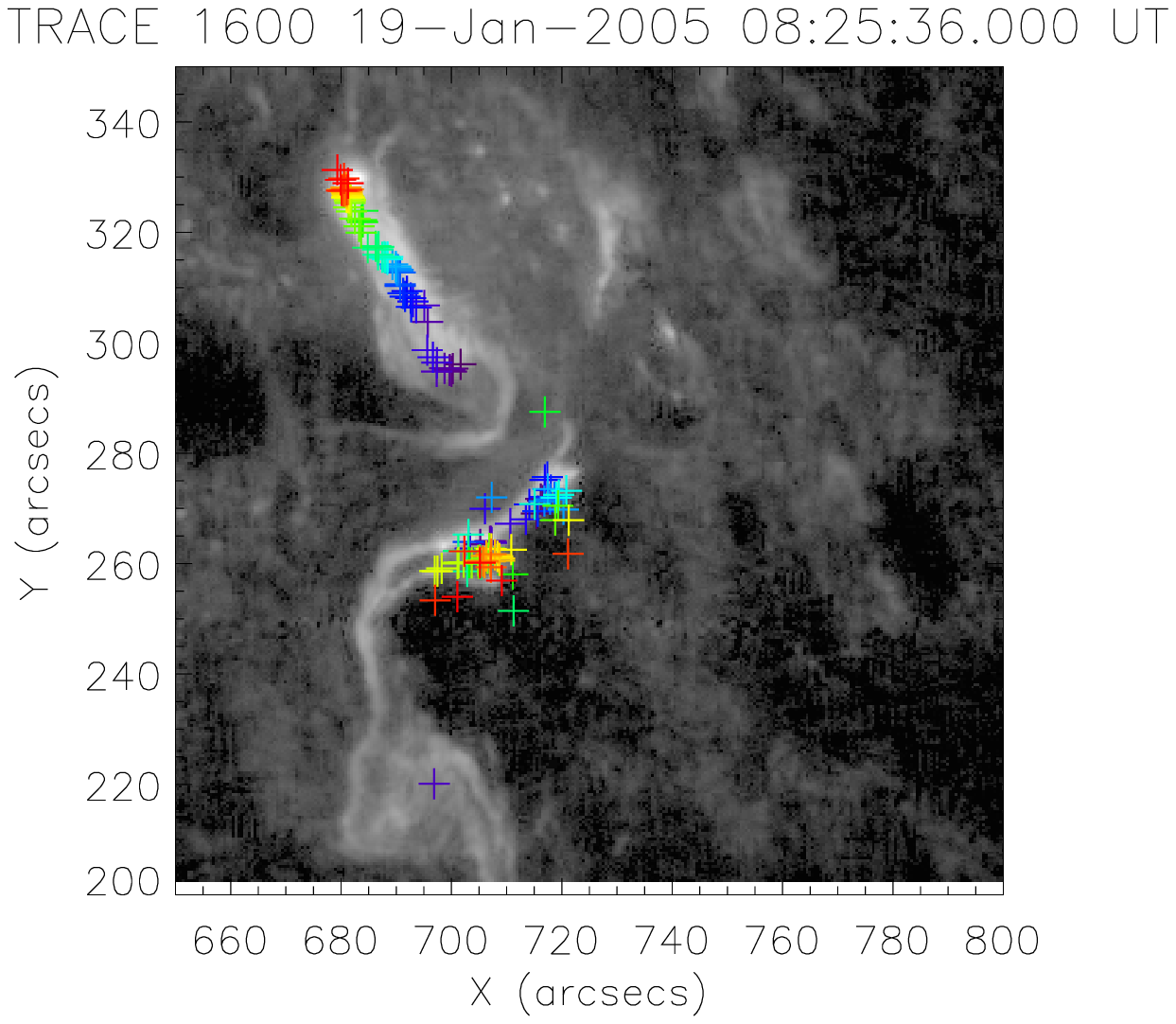}
    \includegraphics[width=8.3cm]{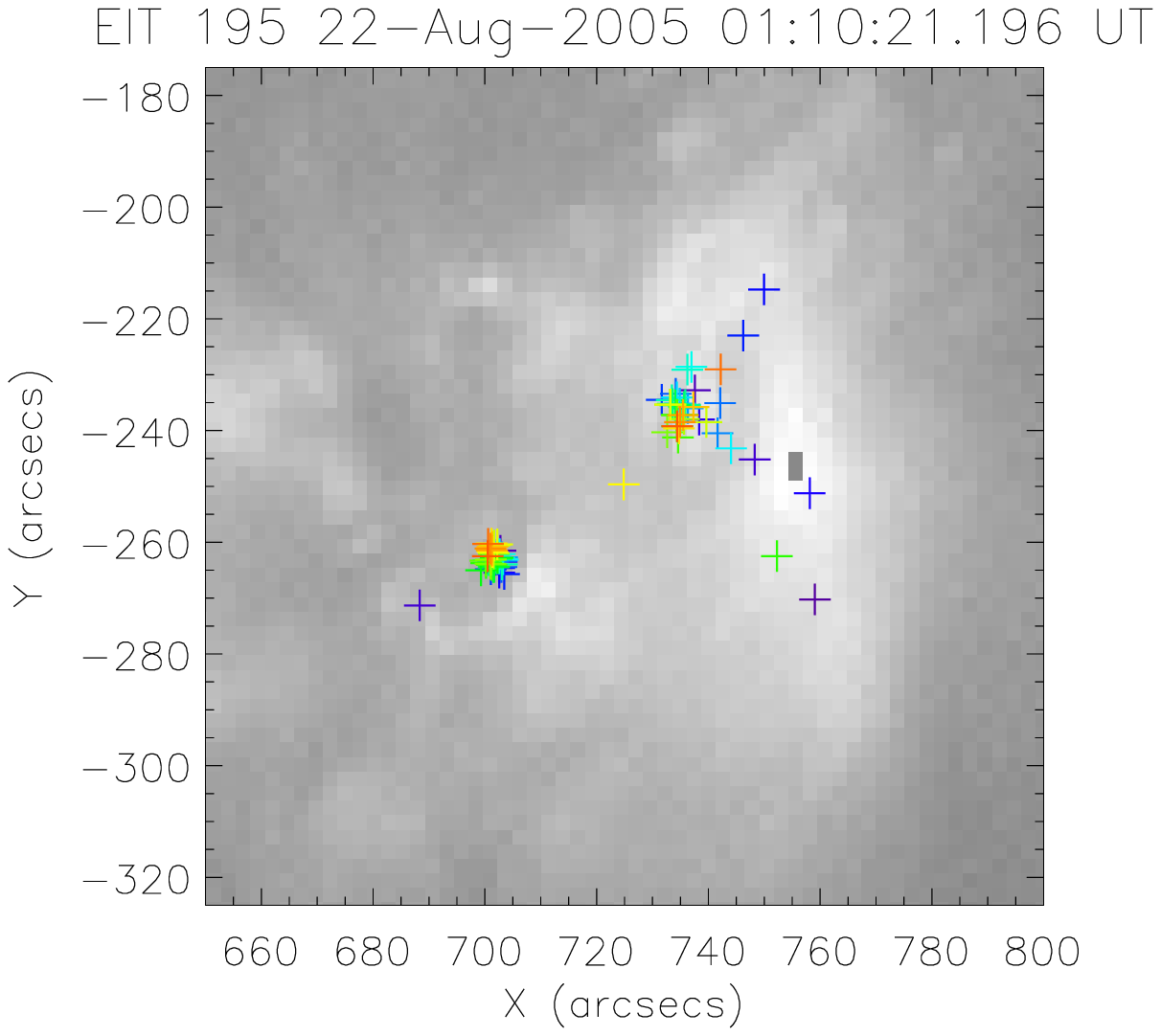}
\caption{Evolution of the locations of the two hard X-ray footpoint sources as a function of time for the three flares. Top panel: The flare of 2002 November 9, with 25 - 50 keV footpoints detected by RHESSI overlaid on a SoHO/EIT image of the flare ribbons. Centre panel: the flare of 2005 January 19, with 35 - 100 keV footpoints overlaid on a TRACE 1600 $\AA$ image. Bottom panel: the flare of 2005 August 22, with 25 - 50 keV footpoints overlaid on a SoHO/EIT image. Time progresses from blue to red in all cases.}
\label{position}
 \end{center}

\end{figure}

Using RHESSI, it is possible to determine the locations from which X-ray emission is occurring. In this study, the focus is on emission in the hard X-ray regime, caused by non-thermal electrons interacting with the chromosphere. Typically this emission takes the form of two footpoint sources. We examine the motion of these footpoints during the three flares. Using the CLEAN algorithm, we reconstruct a sequence of RHESSI images in the hard X-ray energy range. For the 2002 November 9 and 2005 August 22 flares, we utilise the 25-50 keV energy range, while for the 2005 January 19 event we instead use the 35 - 100 keV range, as there remains a significant thermal component to the 25-50 keV emission in this case. Integration times are 12 s for the 2002 November 9 and 2005 January 19 events, while for 2005 August 22 we adopt a 20 s integration time. For each flare, the RHESSI images are manually divided into two regions, each of which contains one footpoint source. The peak position of each source is then found for each image. Images where only one source was evident are discarded. Thus we plot the position of the hard X-ray sources at different times overlaid on EIT or TRACE images taken during the flares, shown in Figure \ref{position}.

%\begin{figure}
% \begin{center}
%  \includegraphics[width=9cm]{xdistance.ps}
% \includegraphics[width=9cm]{ydistance.ps}
%\caption{Evolution in footpoint position in x (top panel) and y(bottom panel) for each HXR source. The left hand footpoint is shown by the solid line, the right hand footpoint by %the dashed line.}
%\label{motion}
% \end{center}

%\end{figure}

For the 2002 November 9 flare the source positions are shown as a function of time superimposed on the SOHO/EIT image of the flare at 13:13 UT, just after the flare onset. Time progresses from blue (at the flare onset at 13:12 UT) to red (at 13:22 UT) - see Figure \ref{position}, top panel. With the absence of TRACE data, imaging of this event in the EUV range is limited to SOHO/EIT observations of 5'' resolution. Nevertheless, there appears to be a double-ribbon structure of significant extent present during this flare. This illustrates that the footpoint sources progress southward along the two ribbons of the flare, converging as they move. The convergence of the footpoint sources is reiterated in Figure \ref{separation} (left).

As a result of these plots, we estimate the mean velocity of each footpoint by combining the motion in x and y. We find that the left hand footpoint moves at $v_{left} \approx 50$ km/s while the right hand footpoint moves at a similar velocity of $v_{right} \approx 45$ km/s. Similar velocities were calculated by \citet{2005ApJ...625L.143G}. As pointed out in \citet{2011ApJ...730L..27N}, such velocities are broadly consistent with the slow wave model.

The 2005 January 19 flare was observed by TRACE in the 1600 $\AA$ bandpass, allowing for finer image resolution and more rapid cadence - 30 s as opposed to 600 s - than is possible with SOHO/EIT. As in the previous event, the hard X-ray footpoint sources of emission are observed to progress significantly along the flare ribbons. On this occasion however, the ribbons are orientated at an angle of approximately 90$^\circ$ to each other, rather than the more acute angle observed in the previous event. In this flare, the footpoint separation increases significantly over the course of the event. However, although the 90$^\circ$ angle suggests that the arcade may be sheared, \citet{2011ApJ...730L..27N} pointed out that the presence of shear should not qualitatively change the expected results of slow wave propagation, hence this is beyond the scope of this paper. The evolution of the hard X-ray footpoints can be seen in the centre panel of Figure \ref{position}, where the peak positions of emission have been located using the same technique as for the previous flare, and time progresses from blue to red as before. However, as the event contains a significant component of thermal emission above 25 keV, we utilise 35-100 keV binned data instead.

%\begin{figure}
% \begin{center}
 % \includegraphics[width=9cm]{19_jan_05_trace_50_100_centroids.ps}
 % \caption{Centroids of 50 - 100 keV emission observed by RHESSI, overlaid on a TRACE 1600 A image.}
 %\end{center}

%\end{figure}

Despite the signature of pulsations in the lightcurves, the sources of the 2005 August 22 flare remain relatively static throughout the event, as shown in the bottom panel of Figure \ref{position}. This is corroborated by measurements performed by \citet{2010ApJ...724..171R}, who noted displacement velocities of $v \approx$ 8 km/s in hard X-ray sources. The right source is not always evident in the images taken during this flare.
%\begin{figure}
% \begin{center}
 % \includegraphics[width=9cm]{peaks_22_aug_05_replace}
 % \caption{Hard X-ray peak flux positions for footpoint sources observed in the 25 - 50 keV energy band as a function of time, overlaid on an EIT 195 A image.}
 % \label{footpoints_22_aug_05}
 %\end{center}

%\end{figure}

%The static nature of the footpoints can be plausibly explained via two scenarios; either a single static loop structure featuring continuous modulated emission, perhaps indicative %of a standing MHD wave, or a series of loop structures rising in height, as is sometimes anticipated in periodic reconnection models.

\section{Footpoint separation and pulse timings}

\begin{figure*}
 \begin{center}
  \includegraphics[width=5.4cm]{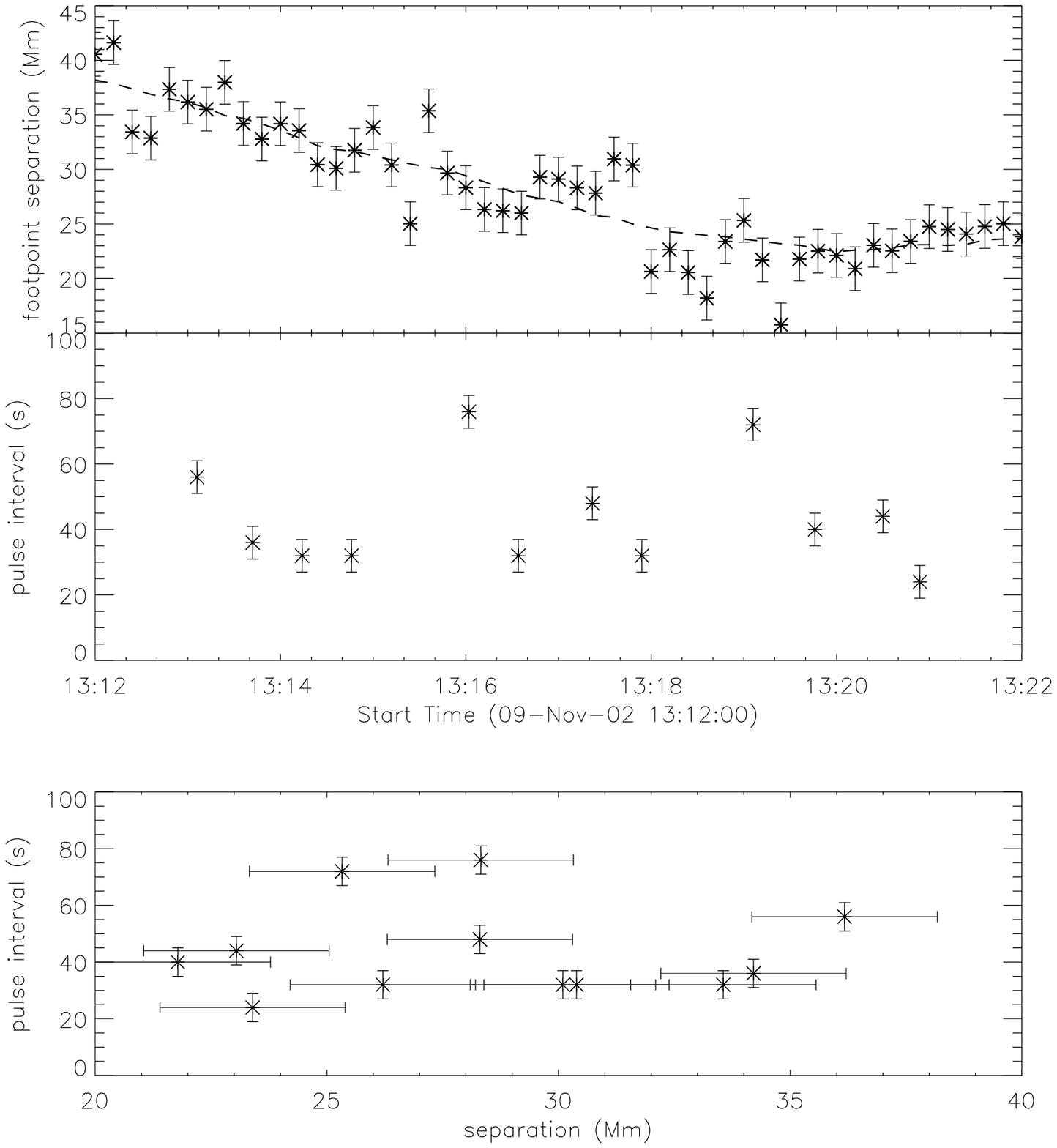}
  \includegraphics[width=5.4cm]{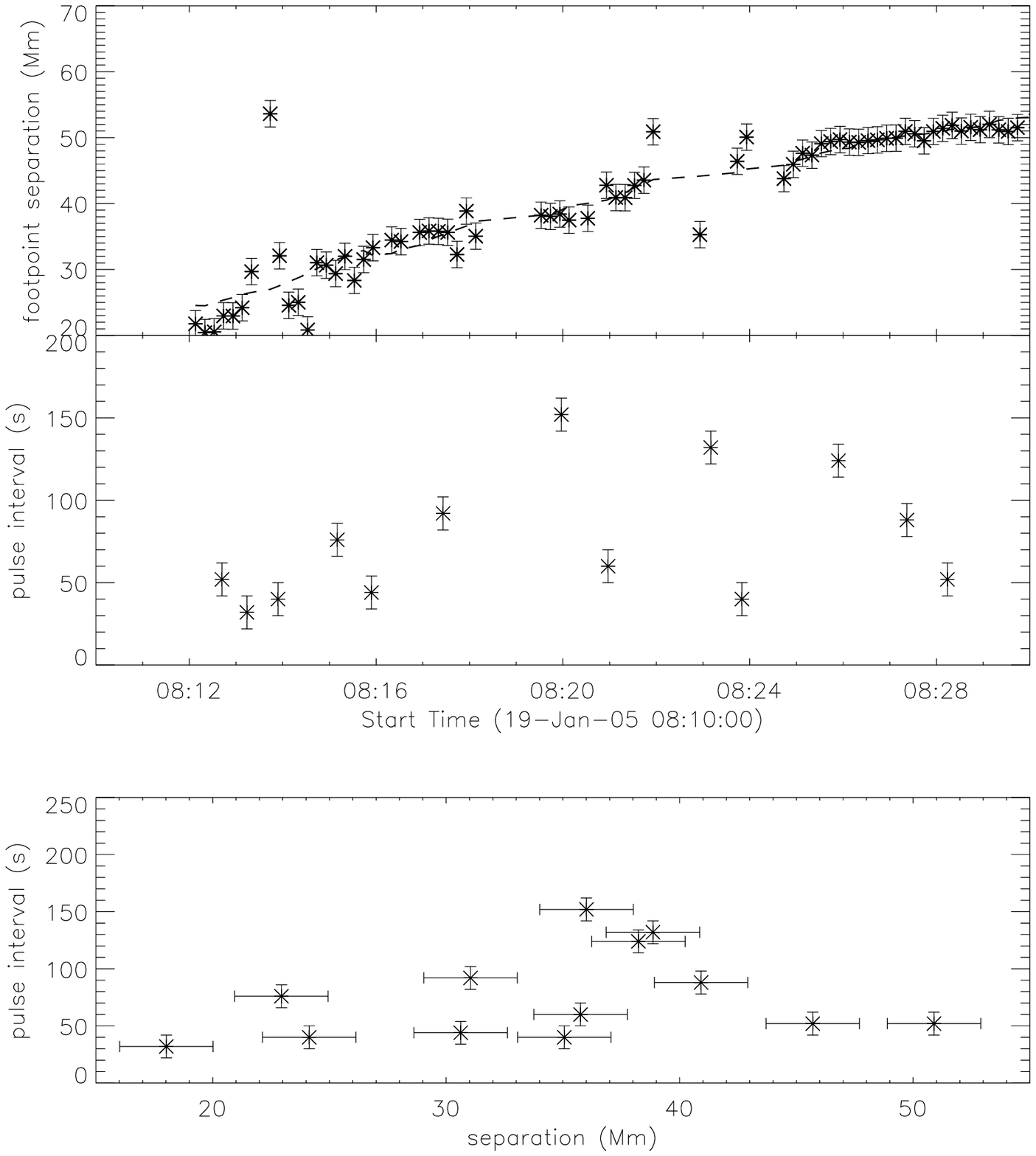}
  \includegraphics[width=5.4cm]{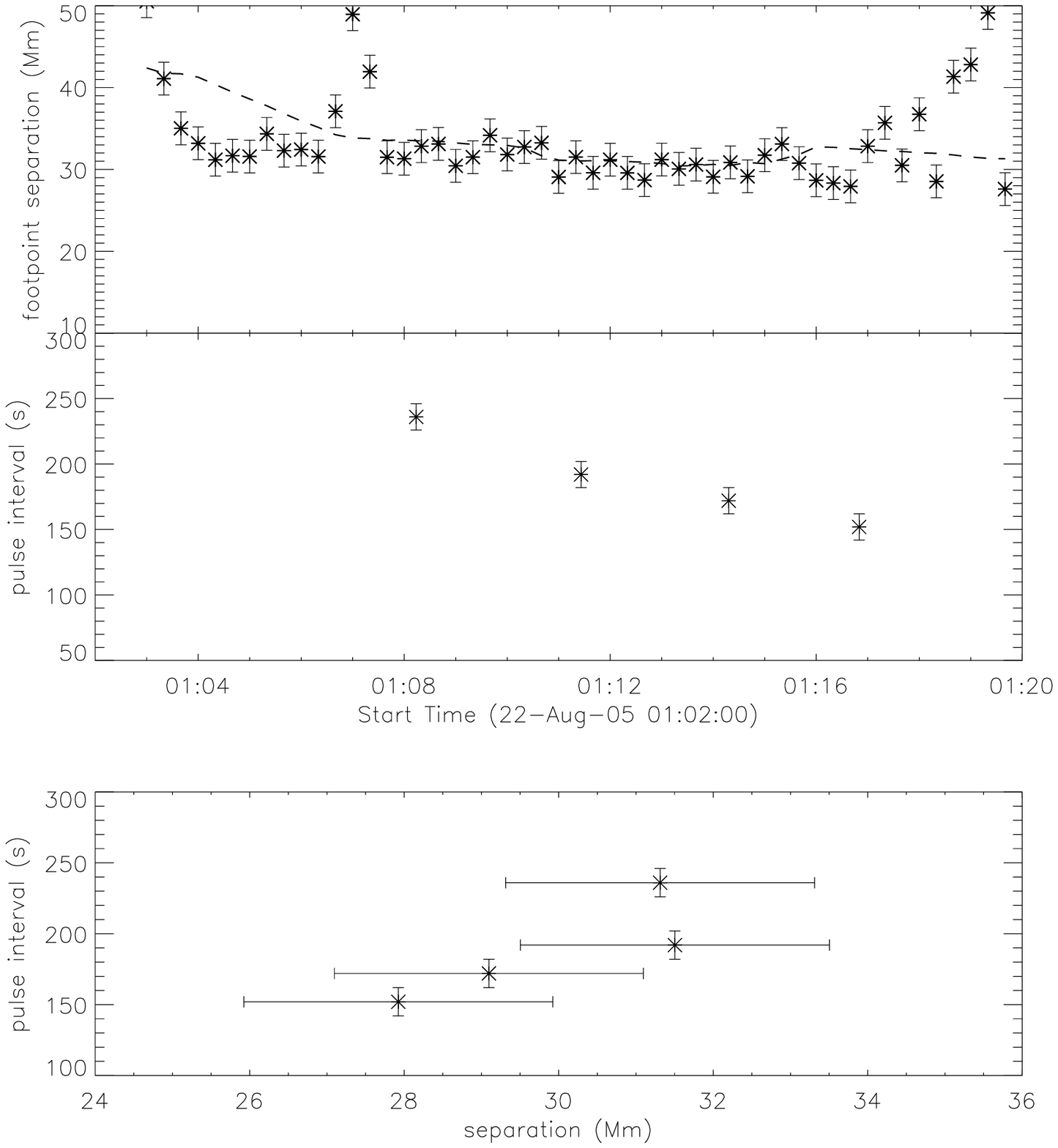}
\caption{Analysis of the relationship between the pulse interval and the footpoint separation for the 2002 November 9, 2005 January 19, and 2005 August 22 flares, from left to right. Top panels: separation of the two hard X-ray footpoints as a function of time. The dashed line is a smoothing of the original data, illustrating the overall trend. Centre panels: time between successive hard X-ray pulses over the same time interval as the top panels. Bottom panels: pulse interval plotted against footpoint separation.}
\label{separation}
 \end{center}

\end{figure*}

Further testing in support of the slow wave model is required, beyond measurements of footpoint velocities. We anticipate that the interval between significant pulses in the lightcurves and the footpoint separation may be correlated. A decrease in the separation of loop footpoints is indicative of a shorter travel distance for the reflected slow waves, and hence the interval between pulses may also be decreased. We estimate the separation of the footpoints for each flare, as shown in the top three panels of Figure \ref{separation}. For the 2002 November 9 flare, a significant decrease is observed as the flare progresses. In contrast, for the 2005 January 19 flare the separation is seen to increase significantly. Finally, in the 2005 August 22 event, the progression of hard X-ray sources is very slow and the separation remains relatively constant throughout the flare.

In Figure \ref{separation}, we plot the footpoint separation and the pulse interval as functions of time for each flare. Also, in the bottom panel for each flare we plot the pulse interval against the corresponding footpoint separation to aid in establishing the presence of any trend. To measure the pulse intervals, we take the time of a pulse and subtract the time at which the previous pulse occurred. The individual pulse times are obtained via a semi-automated search for near-zero values in the lightcurve time derivatives, although in each case the results are verified by eye, as seen in Figure \ref{xray}. For the 2002 November 9 flare, the pulse intervals show no obvious trend to shorter time scales, as we might expect from the contraction of the footpoint separation as the flare progresses. Hence, to first order it is difficult to explain this two-ribbon flare in terms of slow wave propagation, as the pulse timings are not correlated with the footpoint separation. For the 2005 January 19 flare,we observe a similar situation. In this case, the footpoint separation increases substantially as the flare progresses. However, there is no clear corresponding increase in the pulse interval. During the 2005 August 22 event, there is a clear decreasing trend in the pulse interval. This flare also features a relatively simple time series consisting of only five distinct peaks, which are relatively straightforward to localise in time. However, due to the very slow motion of the footpoints, there is a lack of variation in the footpoint separation during the flare, suggesting that the variation in pulse intervals must have some other cause. 

%One difficulty is establishing what to consider as a significant pulse in the lightcurve, as opposed to a random fluctuation. An alternative technique would be to utilise the %wavelet transform to investigate any significant frequencies in the data and how they may change as a function of time. However, for short, impulsive time series data typical for %solar flares, this technique often leads to misleading results due to the difficulty in subtracting an appropriate background and extraneous power at low frequencies, a known %wavelet feature \citep{Torrence98apractical}.

%Repeating our analysis of the 2002 November 9 flare, we investigate the correlation between footpoint separation and the interval between pulses. The results are shown in Figure %\ref{pulses_19_jan_05}.

%\begin{figure}
 %\begin{center}
 % \includegraphics[width=8cm]{19_jan_05_separation.ps}
  %\includegraphics[width=8cm]{pulses_beta_19_jan_05.ps}
  %\caption{Top: separation of the two hard X-ray footpoints in the 50 - 100 keV energy band as a function of time. Bottom: Time between a hard X-ray pulse and its predecessor over %the same time interval.}
%\label{pulses_19_jan_05}
% \end{center}

%\end{figure}
. 
%In this example there is come correlation between the footpoint separation and the timings of hard X-ray pulses; both appear to increase as the flare progresses. A cross %correlation of the two datasets yields a correlation coefficient of [INSERT VALUE]. This suggests that in this case, the loop lengths in the arcade are an important factor %determining the period of the QPP. This, together with the relatively slow footpoint velocities, is consistent with the slow mode interpretation.

%As implied in the previous section, the separation of the hard X-ray footpoints for this flare remains relatively static, as indicated in Figure \ref{separation_22_aug_05}. This %flare also features a relatively simple time series consisting of a few distinct peaks, which are relatively straightforward to localise in time. The pulse intervals are shown in %the bottom panel of Figure \ref{separation_22_aug_05}.

%\begin{figure}
% \begin{center}
 % \includegraphics[width=8cm]{separation_peaks_22_aug_05.ps}
 % \includegraphics[width=8cm]{pulses_22_aug_05_alternate.ps}
 % \caption{Top: separation of hard X-ray footpoint sources in the 25-50 keV energy band as a function of time. Bottom: Time between a hard X-ray pulse and its predecessor over the %same time interval.}
 % \label{separation_22_aug_05}
 %\end{center}

%\end{figure}

\section{Footpoint velocities at lightcurve peaks and valleys}

Another observational feature that is predicted by the slow wave triggering model is that the progression of hard X-ray footpoints is expected to be discontinuous. That is, instead of a smooth progression of the footpoints along the ribbon, the locations of hard X-ray footpoints would jump between locations, as reconnection events are triggered in the arcade at different locations. We can search for this behaviour by examining the apparent velocity of the footpoints as a function of time for each event. According to the slow wave model, we might expect that the apparent footpoint velocity is close to zero during the X-ray peaks, while being non-zero during the intervening periods, although this effect would be mitigated if successive peaks occur closely together in time compared with their characteristic durations. 

\begin{figure}
 \begin{center}
  \includegraphics[width=8cm]{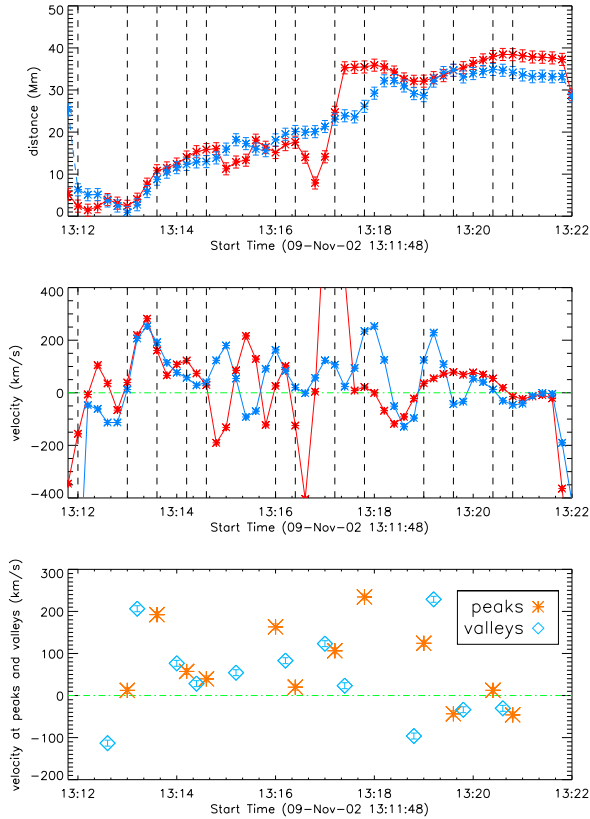}
  \caption{Motion of the hard X-ray footpoints during the solar flare of 2002 November 9. Top panel: The displacement of the left-hand (red) and right-hand (blue) X-ray sources as a function of time. Central panel: The velocity of the footpoints as a function of time. Bottom panel: Footpoint velocity measurements taken for the stronger right source only at each peak (orange) and valley (blue) in the hard X-ray lightcurve observed by RHESSI. In the first two panels the peaks in the hard X-ray lightcurve are represented by vertical lines.}
  \label{derivatives_full}
 \end{center}

\end{figure}

In Figure \ref{derivatives_full} we examine the motion of the footpoints during the 2002 November 9 flare. In this example both footpoints, designated by the red and blue curves in the top panel, are observed to move substantially throughout the course of the flare. The pulsation peaks in the lightcurve are represented by the vertical lines. The central panel shows the velocity of each footpoint as a function of time. In the bottom panel we show only the velocity at each peak and valley location for the left-hand footpoint, denoted by the red line. The results indicate no clear preference for the peak velocities being lower compared to the valley velocities. The velocity is also observed at some peaks and valleys to be negative.

\begin{figure}
 \begin{center}
  \includegraphics[width=8cm]{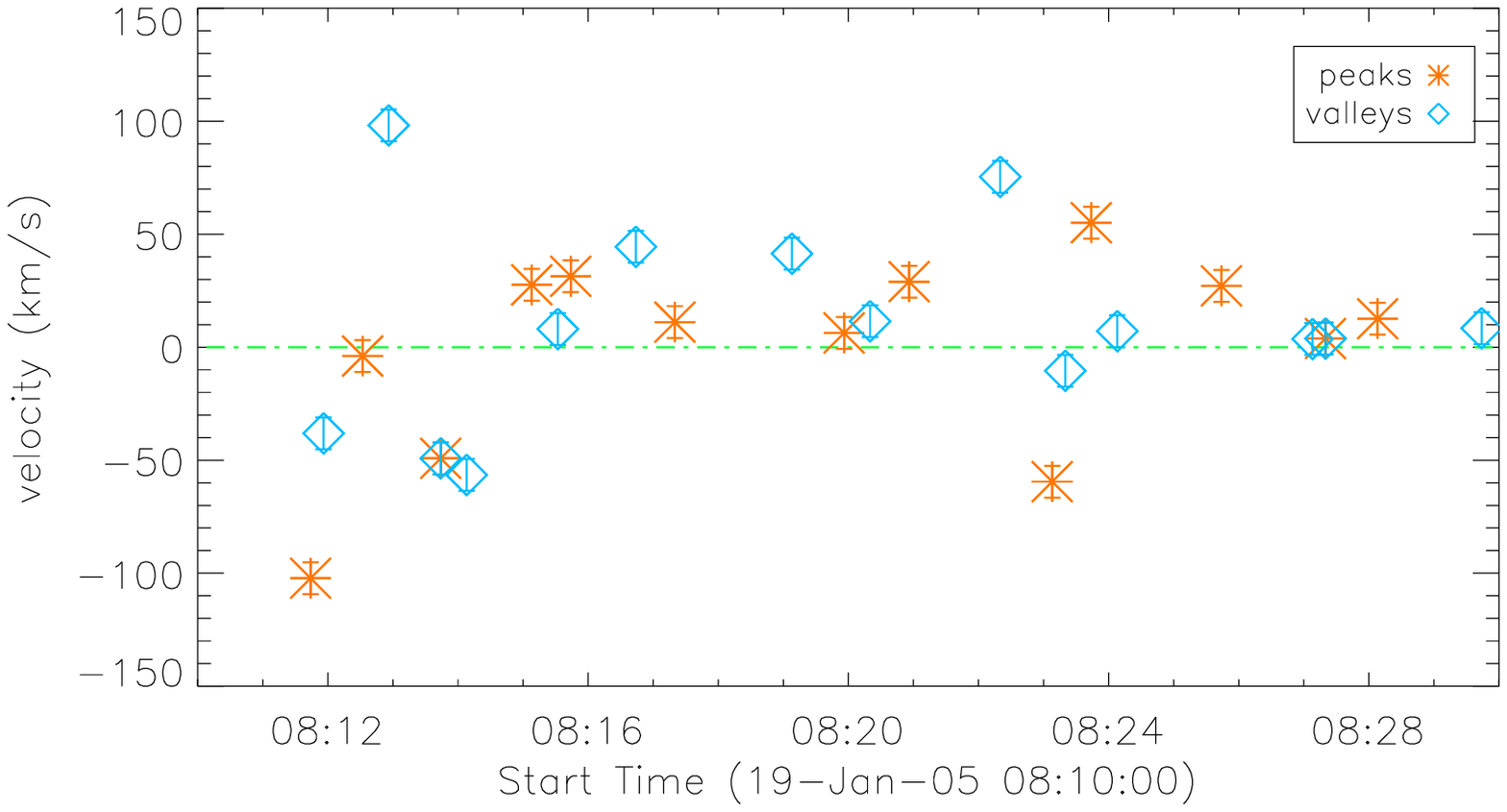}
  \includegraphics[width=8cm]{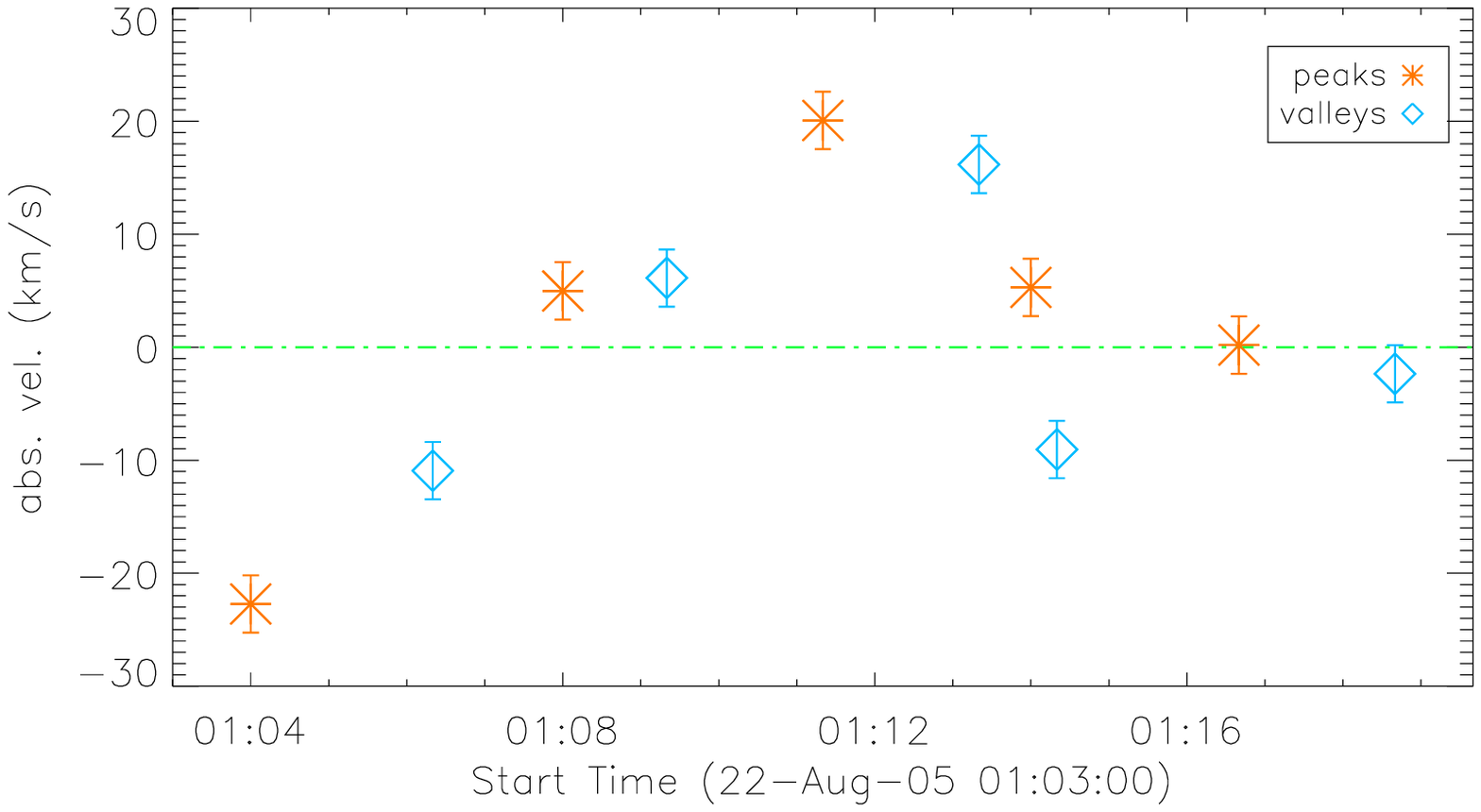}
  \caption{Footpoint velocity measurements taken for the strongest source, in both cases left source, only at each peak (orange) and valley (blue) in the hard X-ray lightcurves observed by RHESSI for the flares of 2005 January 19 (top) and 2005 August 22 (bottom).}
  \label{derivatives_other}
 \end{center}

\end{figure}

The flares of 2005 January 19 and 2005 August 22 are examined in a similar fashion, with the result reproduced in Figure \ref{derivatives_other}. Similarly, the plots show no clear preference for lower velocities at pulse peaks, as opposed to during valleys. Again, at some points the measured instantaneous velocity is negative. Within the uncertainty, the progression of footpoints appears to be smooth, whether in the positive or negative direction, and the velocity is independent of the behaviour of the lightcurve. This is consistent with a continuous motion of the reconnection region in the flaring arcade, where the amount of energy released and therefore hard X-ray production varies in time. This is in contrast with the step-wise footpoint motion that may be expected from the slow wave model. 

\section{Discussion: consistency with a slow wave interpretation}

We have tested whether the observations of three flares are consistent with a reflecting slow wave regime, based on the model proposed by \citet{2011ApJ...730L..27N}. Using measurements of the mean footpoint velocity and the length of arcade loops, it is possible to forward model the expected spatial distance $\delta d_{model}$ between successive hard X-ray maxima and also the expected period $\delta p_{model}$ by assuming a preferential propagation angle of 25 - 28 degrees. It should be noted that, by considering the angle of 25-28 degrees only, we are restricting our attention to the fastest component of the slow waves. In fact, the slow waves would disperse over all smaller angles at slower velocities, as recently illustrated by \citet{2011A&A...536A..68G}. However, \citet{2011ApJ...730L..27N} suggested that the fast component alone would be enough to trigger progressive reconnection events, and it is this hypothesis that we test here.

By modelling the flare arcade in a simple fashion, assuming semi-circular loops, the distance $a$ along the loop from the looptop to the footpoint is given by,

\begin{equation}
 a = \frac{\pi S}{4}
\end{equation}

where $S$ denotes the footpoint separation. However if a slow wave travels at an angle $\theta$ along the arcade as well, we can say that this distance $b$ is approximately given by,

\begin{equation}
 b = \frac{a}{\cos \theta}
\end{equation}

Hence the total travel distance of the wave between recombinations is $2b$, or $2a/\cos \theta$. Meanwhile, the component of distance along the arcade $\delta d_{model}$ is given by,

\begin{equation}
 \delta d_{model} = \frac{\pi S \tan \theta}{2}
 \label{sep_equation}
\end{equation}

The expected pulse period $\delta p_{model}$ is estimated from the measured mean footpoint velocity $v$ and $\delta d_{model}$, via $\delta p_{model} = \delta d_{model} / v$.

For example, during the 2002 November 9 event, we find the footpoint velocity to be approximately $v$ = 50 km/s. Using the footpoint separation to estimate the arcade loop lengths, we can calculate both $\delta d_{model}$ and $\delta p_{model}$. With $S \approx$ 30 Mm, we find that $\delta d_{model} \approx$ 22 - 25 Mm, while $\delta p_{model} \approx$ 440 - 500 s. Hence, although the velocity $v$ is broadly consistent with a slow mode interpretation, the resulting expected distance between hard X-ray peaks is too large to match the observations, and the expected pulse interval is several times greater than the observed $\delta p_{obs}$ = 30 - 80 s. This flare then cannot be easily explained using the slow wave reflection model.

Similar calculations may be made for the other flares studied. During the 2005 January 19 flare, the mean footpoint velocity is in the range 60 km/s \citep[see][]{2009SoPh..258...69Z}. Repeating our calculations, we find that $\delta d_{model} \approx$ 22 - 35 Mm and $\delta p_{model} \approx$ 370 - 600 s. This period is a factor of 3 larger than the measured $\delta p_{obs}$ = 50 - 200 s, while the expected distance between successive hard X-ray maxima remains much too large to be explained by the observed footpoint motions. The total displacement of the hard X-ray footpoints during this event is around 25 - 30 Mm, the distance predicted to occur between just two pulses.

For the 2005 August 22 flare, the velocity of the footpoints is slow. Our measurements concur with those of \citet{2010ApJ...724..171R}, who noted that $v <$ 10 km/s. This leads to $\delta d_{model} \approx$ 22-25 Mm and $\delta p_{model} >$ 2000 s, clearly inconsistent with the observations, where the total displacement of the sources is around 10 Mm and the period is in the region of 100 - 250 s. This information is summarised in Table \ref{table1}.

\begin{table*}
\caption{Summary of flare properties and observed versus predicted parameters for the selected events, in terms of the slow wave reflection model.}
\label{table1}
\centering
\begin{tabular}{c|ccc}
 \tableline
 \tableline
\  & 2002 November 9 & 2005 January 19 & 2005 August 22 \\
%\ Event date & observed $\delta$p (s) & modelled $\delta$p (s) & observed $\delta$d (Mm) & modelled $\delta$d (Mm) \\
\tableline
 GOES start time (UT) & 13:08 & 08:03 & 00:44 \\  
 GOES peak time (UT) & 13:23 & 08:22 & 01:33 \\
 GOES end time (UT) & 13:36 & 08:40 & 02:18 \\
 pulsations onset time (UT) & 13:10 & 08:12 & 01:03 \\
 number of pulses observed & 13 & 14 & 5 \\
 measured $\delta p_{obs}$ (s) & 30 - 80 & 50 - 200 & 100 - 250 \\
 modelled $\delta p_{model}$ (s) & 440 - 500 & 370 - 600 & $>$2000 \\
 measured $\delta d_{obs}$ (Mm) & $<$5 & $<$5 & $<$2  \\
 modelled $\delta d_{model}$ (Mm) & 22 - 25 & 22 - 35 & 22 - 25 \\
 footpoint separation $S$ (Mm) & 20 - 40 & 30 - 50 & 25 - 35 \\
 mean footpoint velocity $v$ (km/s) & 50 - 60 & 50 - 70 & 5 - 10 \\ 
 total footpoint motion $D$ (Mm) & 30 - 40 & 25 - 30 & 5 - 15  \\
\tableline
\end{tabular}

\end{table*}

\begin{figure}[H]
 \begin{center}
  \includegraphics[width=8cm]{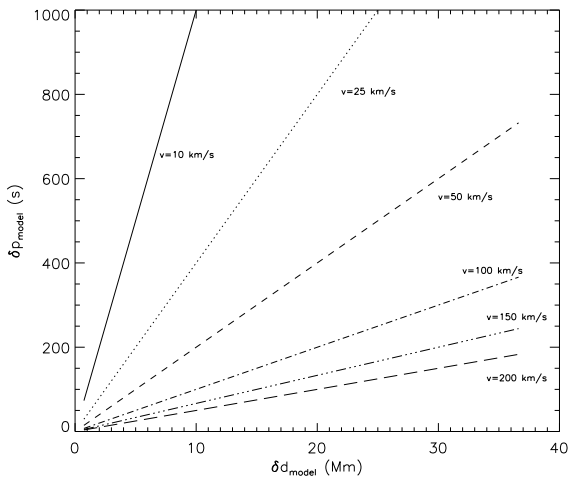}
  \caption{Parameterisation of the observational characteristics expected of slow waves in two-ribbon flares. For varying values of $v$, we estimate the expected values of $\delta d_{model}$ and $\delta p_{model}$ in the arcade. The footpoint separation $S$ is directly related to $\delta d$ via Equation \ref{sep_equation}. }
 \label{parameters}
 \end{center}

\end{figure}

It was also shown in Figure \ref{derivatives_full} and Figure \ref{derivatives_other} that there was no preference for $v$=0 in the vicinity of lightcurve peaks, as opposed to lightcurve valleys. Such a preference would have been suggestive of the slow wave reflection model, where, although the slow waves travel continuously, they only superimpose and trigger magnetic reconnection at discrete locations in the arcade's coronal plasma. Hence, the production of accelerated electrons which travel to the chromosphere would also occur discretely, and as a result the motion of hard X-ray footpoints would not be constant in time. Instead, the observations suggest a relatively smooth progression of footpoints as a function of time, independent of lightcurve behaviour, although we emphasise that the uncertainty is considerable.

We conclude that these three flares cannot be easily explained by the slow wave reflection model when considering the fastest component of the slow wave distribution.  The main reason for this is the expected value of the propagation angle $\theta$. Utilising a preferential reflection angle of 25-28 degrees, the expected distances between successive hard X-ray maxima are much larger than the observed values. Typically the total movement of the hard X-ray sources between pulses is small, less than 10 Mm. The observed periods are also much shorter than those predicted, by a factor of 5-10. The 2005 August 22 flare could instead be considered in the context of slow waves reflecting directly back along field lines, as the total displacement of the sources is so small.

As a guide, we parameterise the problem in Figure \ref{parameters}, illustrating the expected values of $\delta d$ and $\delta p$ from a range of values of footpoint velocity $v$ and footpoint separation $S$. In general, it appears that pulsations with $\delta p$ less than $\approx$ 60 s cannot be realistically explained by this model, given our expectations of slow wave velocities being in the range 100 - 300 km/s. Such short periods would require either ribbons extremely close together, or a very fast, unrealistic motion of the footpoints along the ribbons. However, longer period oscillations may still be consistent with a slow wave regime. For example, an observed period of $\delta p$ = 300 s, coupled with a ribbon separation of $S$ = 10 Mm, would lead to the estimates $\delta d \approx$ 8 Mm and $v \approx$ 25 km/s, which are within the range of previous observations. From Figure \ref{parameters}, we suggest that future searches for evidence of slow wave triggering in two-ribbon flares should focus on events where either $\delta p$ is large and $v$ is small, or \textit{vice versa}.

Although we find it difficult to explain these three events in terms of the slow-wave theory, an important consideration is that the fastest component of the slow waves may not carry enough energy to trigger the progressive bursts of reconnection. An alternative possibility is that the bulk of the wave energy is travelling more slowly and at a more acute angle than we have investigated here. If these slower waves act as the reconnection trigger, the observations would become more consistent with the model. Hence, our predictions of observable parameters based on the 25-28 degree angle should be considered as upper limits on these parameters.

In making these estimates we have considered a first-order, idealised 2-dimensional scenario, ignoring any non-uniformity in loop height and assuming that the arcade loops may be considered by assuming a semi-circle between footpoints and ignoring additional curvature effects. Considering a full 3-dimensional model would undoubtedly have some effect on the estimates presented here. For example, a more realistic arcade loop may be deformed and stretch higher into the corona than a simple semi-circular model, resulting in a longer travel path for slow waves. We also neglect the presence of shear or twist in the arcade loops for simplicity. However, these estimates still provide a useful first approximation as to whether pulsations in the flares studied here can be realistically explained via slow waves. Further study of such two-ribbon flares is needed to understand whether other examples of quasi-periodic pulsations may be explained in this way.

\begin{acknowledgements}
 We are grateful to Anil Gopie, Nicholas Shields and Richard Schwartz for helpful discussions on the topic of peak finding in lightcurves. ARI is also grateful to Valery Nakariakov for useful discussions on the propagation of slow waves in flares. ARI was supported by an appointment to the NASA Postdoctoral Program at Goddard Space Flight Center, administered by Oak Ridge Associated Universities through a contract with NASA. 
\end{acknowledgements}

\bibliographystyle{aa}
\bibliography{pulse_timings}

\begin{thebibliography}{33}
\expandafter\ifx\csname natexlab\endcsname\relax\def\natexlab#1{#1}\fi

\bibitem[{{Asai} {et~al.}(2001){Asai}, {Shimojo}, {Isobe}, {Morimoto},
  {Yokoyama}, {Shibasaki}, \& {Nakajima}}]{2001ApJ...562L.103A}
{Asai}, A., {Shimojo}, M., {Isobe}, H., {et~al.} 2001, \apjl, 562, L103

\bibitem[{{Chiu}(1970)}]{1970SoPh...13..420C}
{Chiu}, Y.~T. 1970, \solphys, 13, 420

\bibitem[{{Edwin} \& {Roberts}(1982)}]{1982SoPh...76..239E}
{Edwin}, P.~M. \& {Roberts}, B. 1982, \solphys, 76, 239

\bibitem[{{Edwin} \& {Roberts}(1983)}]{1983SoPh...88..179E}
{Edwin}, P.~M. \& {Roberts}, B. 1983, \solphys, 88, 179

\bibitem[{{Fletcher} {et~al.}(2011){Fletcher}, {Dennis}, {Hudson}, {Krucker},
  {Phillips}, {Veronig}, {Battaglia}, {Bone}, {Caspi}, {Chen}, {Gallagher},
  {Grigis}, {Ji}, {Liu}, {Milligan}, \& {Temmer}}]{2011SSRv..159...19F}
{Fletcher}, L., {Dennis}, B.~R., {Hudson}, H.~S., {et~al.} 2011, \ssr, 159, 19

\bibitem[{{Foullon} {et~al.}(2005){Foullon}, {Verwichte}, {Nakariakov}, \&
  {Fletcher}}]{2005A&A...440L..59F}
{Foullon}, C., {Verwichte}, E., {Nakariakov}, V.~M., \& {Fletcher}, L. 2005,
  \aap, 440, L59

\bibitem[{Grechnev {et~al.}(2003)Grechnev, White, \& Kundu}]{Grechnev}
Grechnev, V.~V., White, S.~M., \& Kundu, M.~R. 2003, \apj, 588, 1163

\bibitem[{{Grigis} \& {Benz}(2005)}]{2005ApJ...625L.143G}
{Grigis}, P.~C. \& {Benz}, A.~O. 2005, \apjl, 625, L143

\bibitem[{{Gruber} {et~al.}(2011){Gruber}, {Lachowicz}, {Bissaldi}, {Briggs},
  {Connaughton}, {Greiner}, {van der Horst}, {Kanbach}, {Rau}, {Bhat}, {Diehl},
  {von Kienlin}, {Kippen}, {Meegan}, {Paciesas}, {Preece}, \&
  {Wilson-Hodge}}]{2011A&A...533A..61G}
{Gruber}, D., {Lachowicz}, P., {Bissaldi}, E., {et~al.} 2011, \aap, 533, A61+

\bibitem[{{Gruszecki} \& {Nakariakov}(2011)}]{2011A&A...536A..68G}
{Gruszecki}, M. \& {Nakariakov}, V.~M. 2011, \aap, 536, A68

\bibitem[{{Inglis} \& {Nakariakov}(2009)}]{2009A&A...493..259I}
{Inglis}, A.~R. \& {Nakariakov}, V.~M. 2009, \aap, 493, 259

\bibitem[{{Inglis} {et~al.}(2008){Inglis}, {Nakariakov}, \&
  {Melnikov}}]{2008A&A...487.1147I}
{Inglis}, A.~R., {Nakariakov}, V.~M., \& {Melnikov}, V.~F. 2008, \aap, 487,
  1147

\bibitem[{{Inglis} {et~al.}(2011){Inglis}, {Zimovets}, {Dennis}, {Kontar},
  {Nakariakov}, {Struminsky}, \& {Tolbert}}]{2011A&A...530A..47I}
{Inglis}, A.~R., {Zimovets}, I.~V., {Dennis}, B.~R., {et~al.} 2011, \aap, 530,
  A47+

\bibitem[{{Kane} {et~al.}(1983){Kane}, {Kai}, {Kosugi}, {Enome}, {Landecker},
  \& {McKenzie}}]{1983ApJ...271..376K}
{Kane}, S.~R., {Kai}, K., {Kosugi}, T., {et~al.} 1983, \apj, 271, 376

\bibitem[{{Kaufmann} {et~al.}(2009){Kaufmann}, {Gim{\'e}nez de Castro},
  {Correia}, {Costa}, {Raulin}, \& {V{\'a}lio}}]{2009ApJ...697..420K}
{Kaufmann}, P., {Gim{\'e}nez de Castro}, C.~G., {Correia}, E., {et~al.} 2009,
  \apj, 697, 420

\bibitem[{{Li} \& {Gan}(2008)}]{2008SoPh..247...77L}
{Li}, Y.~P. \& {Gan}, W.~Q. 2008, \solphys, 247, 77

\bibitem[{{Lin} {et~al.}(2002){Lin}, {Dennis}, {Hurford}, {Smith}, {Zehnder},
  {Harvey}, {Curtis}, {Pankow}, {Turin}, {Bester}, {Csillaghy}, {Lewis},
  {Madden}, {van Beek}, {Appleby}, {Raudorf}, {McTiernan}, {Ramaty}, {Schmahl},
  {Schwartz}, {Krucker}, {Abiad}, {Quinn}, {Berg}, {Hashii}, {Sterling},
  {Jackson}, {Pratt}, {Campbell}, {Malone}, {Landis}, {Barrington-Leigh},
  {Slassi-Sennou}, {Cork}, {Clark}, {Amato}, {Orwig}, {Boyle}, {Banks},
  {Shirey}, {Tolbert}, {Zarro}, {Snow}, {Thomsen}, {Henneck}, {McHedlishvili},
  {Ming}, {Fivian}, {Jordan}, {Wanner}, {Crubb}, {Preble}, {Matranga}, {Benz},
  {Hudson}, {Canfield}, {Holman}, {Crannell}, {Kosugi}, {Emslie}, {Vilmer},
  {Brown}, {Johns-Krull}, {Aschwanden}, {Metcalf}, \&
  {Conway}}]{2002SoPh..210....3L}
{Lin}, R.~P., {Dennis}, B.~R., {Hurford}, G.~J., {et~al.} 2002, \solphys, 210,
  3

\bibitem[{{Linton} \& {Longcope}(2006)}]{2006ApJ...642.1177L}
{Linton}, M.~G. \& {Longcope}, D.~W. 2006, \apj, 642, 1177

\bibitem[{{Mariska}(2005)}]{2005ApJ...620L..67M}
{Mariska}, J.~T. 2005, \apjl, 620, L67

\bibitem[{{McLaughlin} {et~al.}(2009){McLaughlin}, {De Moortel}, {Hood}, \&
  {Brady}}]{2009A&A...493..227M}
{McLaughlin}, J.~A., {De Moortel}, I., {Hood}, A.~W., \& {Brady}, C.~S. 2009,
  \aap, 493, 227

\bibitem[{Melnikov {et~al.}(2005)Melnikov, Reznikova, Shibasaki, \&
  Nakariakov}]{Melnikov}
Melnikov, V.~F., Reznikova, V.~E., Shibasaki, K., \& Nakariakov, V.~M. 2005,
  \aap, 439, 727

\bibitem[{{Murray} {et~al.}(2009){Murray}, {van Driel-Gesztelyi}, \&
  {Baker}}]{2009A&A...494..329M}
{Murray}, M.~J., {van Driel-Gesztelyi}, L., \& {Baker}, D. 2009, \aap, 494, 329

\bibitem[{{Nakajima} {et~al.}(1994){Nakajima}, {Nishio}, {Enome}, {Shibasaki},
  {Takano}, {Hanaoka}, {Torii}, {Sekiguchi}, {Bushimata}, {Kawashima},
  {Shinohara}, {Irimajiri}, {Koshiishi}, {Kosugi}, {Shiomi}, {Sawa}, \&
  {Kai}}]{1994IEEEP..82..705N}
{Nakajima}, H., {Nishio}, M., {Enome}, S., {et~al.} 1994, IEEE Proceedings, 82,
  705

\bibitem[{{Nakariakov} {et~al.}(2010){Nakariakov}, {Foullon}, {Myagkova}, \&
  {Inglis}}]{2010ApJ...708L..47N}
{Nakariakov}, V.~M., {Foullon}, C., {Myagkova}, I.~N., \& {Inglis}, A.~R. 2010,
  \apjl, 708, L47

\bibitem[{{Nakariakov} \& {Melnikov}(2009)}]{2009SSRv..149..119N}
{Nakariakov}, V.~M. \& {Melnikov}, V.~F. 2009, \ssr, 149, 119

\bibitem[{{Nakariakov} \& {Verwichte}(2005)}]{2005LRSP....2....3N}
{Nakariakov}, V.~M. \& {Verwichte}, E. 2005, Living Reviews in Solar Physics,
  2, 3

\bibitem[{{Nakariakov} \& {Zimovets}(2011)}]{2011ApJ...730L..27N}
{Nakariakov}, V.~M. \& {Zimovets}, I.~V. 2011, \apjl, 730, L27+

\bibitem[{{Parks} \& {Winckler}(1969)}]{1969ApJ...155L.117P}
{Parks}, G.~K. \& {Winckler}, J.~R. 1969, \apjl, 155, L117+

\bibitem[{{Reznikova} {et~al.}(2010){Reznikova}, {Melnikov}, {Ji}, \&
  {Shibasaki}}]{2010ApJ...724..171R}
{Reznikova}, V.~E., {Melnikov}, V.~F., {Ji}, H., \& {Shibasaki}, K. 2010, \apj,
  724, 171

\bibitem[{{Reznikova} \& {Shibasaki}(2011)}]{2011A&A...525A.112R}
{Reznikova}, V.~E. \& {Shibasaki}, K. 2011, \aap, 525, A112

\bibitem[{{Srivastava} {et~al.}(2008){Srivastava}, {Zaqarashvili}, {Uddin},
  {Dwivedi}, \& {Kumar}}]{2008MNRAS.388.1899S}
{Srivastava}, A.~K., {Zaqarashvili}, T.~V., {Uddin}, W., {Dwivedi}, B.~N., \&
  {Kumar}, P. 2008, \mnras, 388, 1899

\bibitem[{{Yang} {et~al.}(2009){Yang}, {Cheng}, {Krucker}, {Lin}, \&
  {Ip}}]{2009ApJ...693..132Y}
{Yang}, Y.-H., {Cheng}, C.~Z., {Krucker}, S., {Lin}, R.~P., \& {Ip}, W.~H.
  2009, \apj, 693, 132

\bibitem[{{Zimovets} \& {Struminsky}(2009)}]{2009SoPh..258...69Z}
{Zimovets}, I.~V. \& {Struminsky}, A.~B. 2009, \solphys, 258, 69

\end{thebibliography}

\end{document}